\documentclass[
    ,draft            % use draft while you are working on the paper
  ]
  {aipproc}

\layoutstyle{6x9}

\begin{document}

\title{Hydrogen issue in Core Collapse Supernovae}
%shape and shift of, 32.70.Jz, 33.70.Jg

%<Replace this text with PACS numbers; choose from this list:
%                \texttt{http://www.aip..org/pacs/index.html}>}

\author{A. Elmhamdi}{
  address={Osservatorio Astronomico di Trieste, Via G.B.Tiepolo 11 - I-34131
	 Trieste - Italy }
}
\author{I.J. Danziger}{
 address={<common address for author1 and author2>}
}
\author{D. Branch}{
  address={ Department of Physics and Astronomy, University of Oklahoma, Norman, OK 73019, USA }
}
\author{B. Leibundgut}{
  address={ European Southern Observatory, Karl-Schwarzschild-Strasse 2. D-85478, Graching - Germany}
}

\begin{abstract}
 We discuss results of analyzing a time series of selected 
 photospheric-optical 
 spectra of core collapse supernovae (CCSNe). This is accomplished by 
 means of the parameterized supernovae synthetic spectrum (SSp) code 
 ``SYNOW''. 
 Special attention is addressed to traces of hydrogen at early phases, 
 especially for the stripped-envelope SNe (i.e. SNe Ib-c). A thin low mass 
 hydrogen layer extending to very high ejection velocities above the helium 
 shell, is found to be the most likely scenario for Type Ib SNe. 

\end{abstract}

\maketitle

%%%%%%%%%%%%%%%%%%%%%%%%%%%%%%%%%%%%%%%%%%%%
%% MAINMATTER
%%%%%%%%%%%%%%%%%%%%%%%%%%%%%%%%%%%%%%%%%%%%

\section{Introduction}
 Our main goals are to identify traces of hydrogen, helium and
 oxygen in type Ib-c SNe, and to identify any systematic similarities 
 and differences correlating with other physical properties in the 
 CCSNe family.
  
 We have selected a sample of 45 photospheric spectra of 20 CCSNe objects.
 In the present paper we show results for a few objects. A detailed 
 analysis for the complete sample has been presented recently 
 in Elmhamdi et al. 2006 \cite{Elm06}.
 For the purpose of our analysis we make use of the parameterized supernova 
 synthetic spectrum code ``SYNOW''. The main task is to identify lines and
 put constraints on the velocities. 
 SYNOW has a number of free input fitting parameters. The most important are: 
 1. ($\tau _{ref}$), the optical depth of ``the reference line''. 2. 
 ($T_{bb}$): the underlying blackbody continuum temperature.
 3. ($v_{phot}$): the velocity at the photosphere, estimated  from 
 the match with Fe II lines.
 The radial dependence of the line optical depths can be chosen to be 
 either exponential with an $e$-folding velocity ``$v_e$'' as free 
 parameter (i.e. $\tau \propto exp(-v/v_e)$), or a power-law with an 
 index ``$n$'' (i.e. $\tau \propto v^{-n}$). 
 Detailed description of the code can be found elsewhere
 \cite{Jeff90,Br02,Elm06}

 An important aspect to understand within the ``SYNOW'' line formation
 logic is the ``detachment'' concept. When assigning a minimum ion velocity, 
 $v_{min}$, greater than the photospheric one 
 $v_{phot}$, the ion is said to be ``detached'' from the photosphere, and 
 consequently has a ``non-zero'' optical depth only starting at $v_{min}$. 
 In ``SYNOW'' the profile of a detached line has a flat-topped
 emission, and the absorption minimum is blueshifted by the detachment
 velocity. An undetached line has a rounded emission peak. A slightly
 detached line has a flat top but only over a small wavelength interval.
 These possibilities are illustrated in Fig. 1(left), where the code run 
 adopts $v_{phot}$=10000 km s$^{-1}$ and T$_{bb}$=6000 K,
  including only hydrogen Balmer lines with $\tau$(H$\alpha)=10$ and 
 having 3 different detachment velocities.   

 We introduce the parameter ``contrast velocity'', defined as: 
 $v_{cont}(line)=v_{min}(line)-v_{phot}$. We discuss as well the behaviour 
 of a similar parameter, defined instead as a ratio, i.e. 
 $v_{cont}^{ratio}(line)=v_{min}(line)/v_{phot}$.
%%%%%%%%%%%%%%%%%%%%%%%%%%%%%%%%%%
\begin{figure}
  \includegraphics[height=.3\textheight]{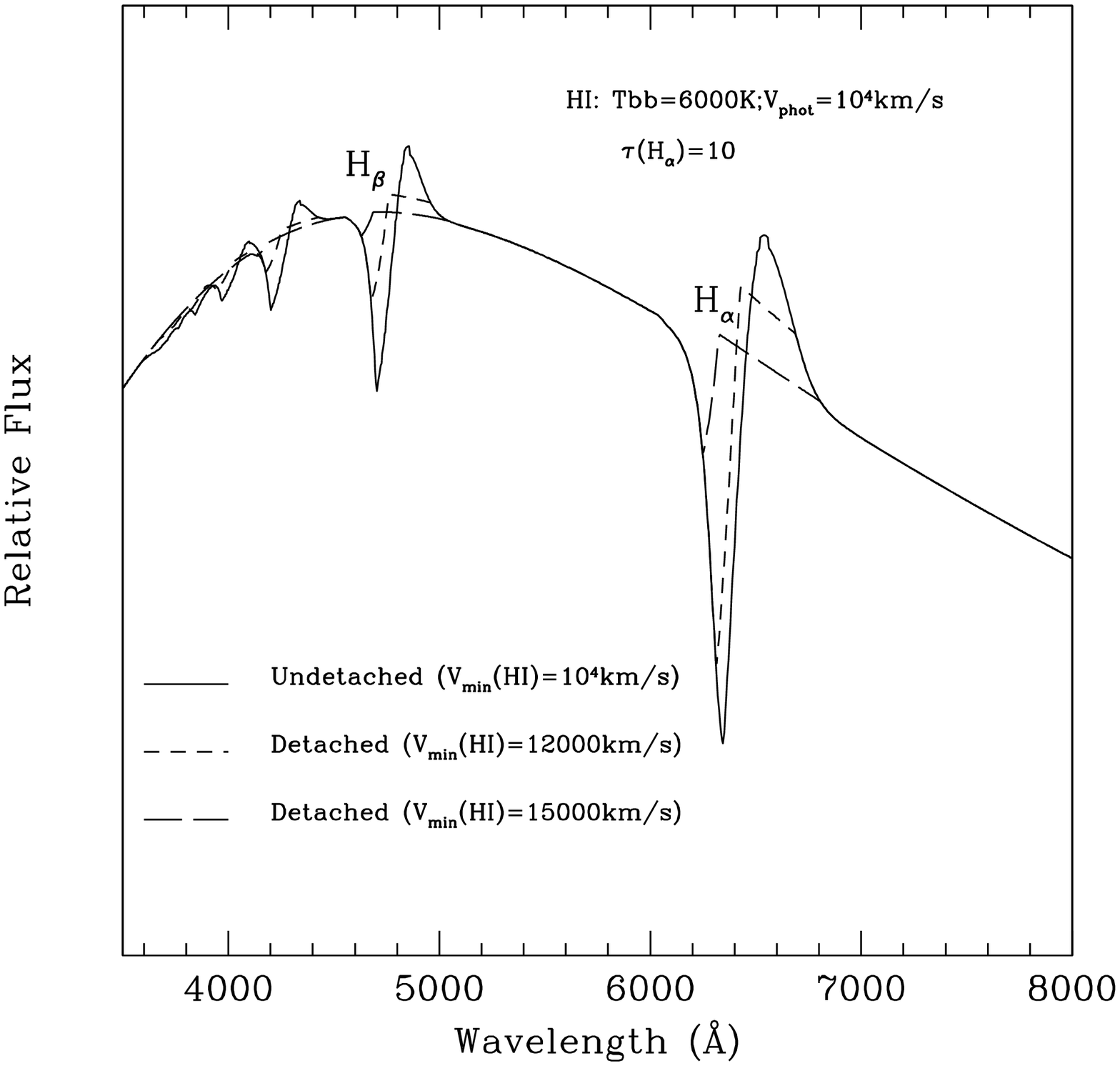}\includegraphics[height=.3\textheight]{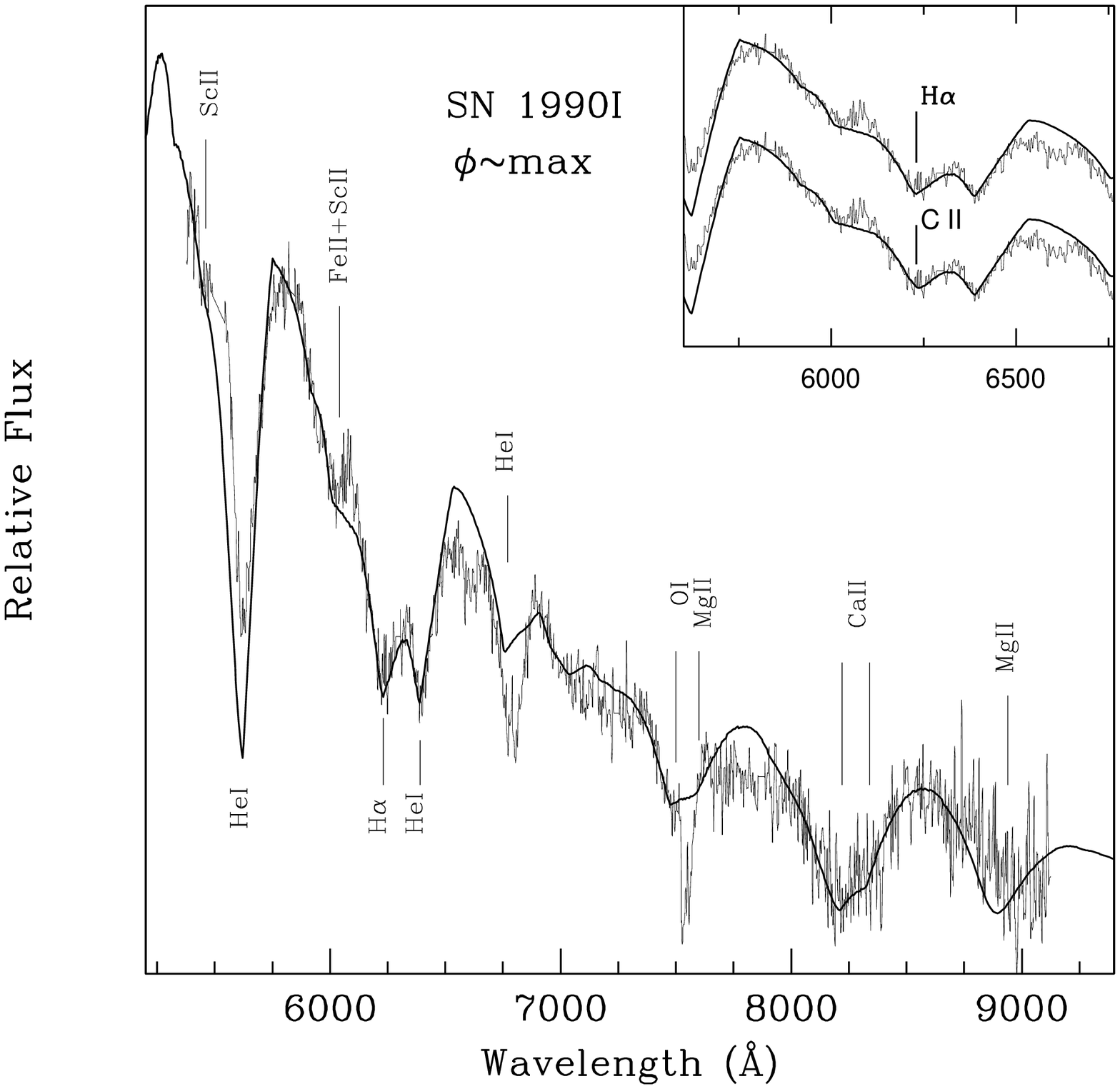}
  \caption{$Left$: the ``SYNOW'' run for Balmer H I lines, using 
  $v_{phot}$=10000 km s$^{-1}$ and T$_{bb}$=6000 K. The resulted profiles 
  for different detachment velocities are illustrated. $Right$: The SSp fit,
  thin line,
  to the spectrum of SN 1990I near maximum. The region around the 6250\AA~
  trough is zoomed in the window.}
\end{figure}
%%%%%%%%%%%%%%%%%%%%%%%%%%%%%%%%%%
\section{Hydrogen issue in CCSNe}
 In the following we present results for some individual objects
 of the CCSNe sample. 

 {\bf SN 1990I :} the event is one of the well sampled 
 and exploited, both photometrically and spectroscopically, objects 
 among type Ib-c SNe\cite{Elm04}.
 Fig. 1(right) compares the observed spectrum at maximum light to a 
 synthetic spectrum (SSp) that has a velocity at the photosphere 
 $v_{phot} = 12000$ km s$^{-1}$ and T$_{bb}=$14000 K. The introduced lines are 
 He I, Fe II, Sc II, Mg II, O I, Ca II and H$\alpha$.
 Absorption troughs of P-Cygni He I lines at 5876\AA, 
 6678\AA~ and 7056\AA~ are evident, $\tau$(HeI)$\sim 2.9$ and $v_{min}$(HeI)$
 =14000$ km s$^{-1}$, although they are not simultaneously 
 fitted (i.e. their relative strengths). This
 limitation is faced each
 time we analyze and fit He I lines using the LTE approach. 
 A more precise analysis requires NLTE treatment as He I lines are probably 
 non-thermally excited by the decay products of $^{56}$Ni and $^{56}$Co 
 \cite{Luc91}. Apart from He I and H$\alpha$ lines, the remaining lines are
 undetached. 
 The absorption minimum near 6250\AA~is well fitted by H$\alpha$, with 
 $v_{min}$(H$\alpha$)=16000 km s$^{-1}$ and assigned a moderate optical depth 
 of 0.6. We tested however other plausible alternative identifications 
  to account for a similar feature seen 
 in type Ib SNe, namely Si II 6355\AA, Ne I 6402\AA~and C II 6580\AA~lines.
 For SN 1990I, $v_{cont}$( H$\alpha$) is only 4000~km~s$^{-1}$, hence
  undetached Ne I line is excluded because it is too blue to fit the observed 
 feature. Si II 6355\AA~is obviously also too blue. C II 6580\AA~remains then 
 a plausible alternative for H$\alpha$, since its rest wavelength is slightly 
 redder than H$\alpha$ and therefore has its contrast velocity as a free 
 parameter. In the window of Fig. 1(right), we show the H$\alpha$ and C II fit
 cases. C II, with 
 $v_{min}$(CII)=17000 km s$^{-1}$ and
 $\tau$(CII)$\sim 0.003$, provides a fit as good as H$\alpha$. But an 
 acceptable C II 6580\AA~line would require an 
 additional velocity of about 820 km s$^{-1}$  compared to H$\alpha$ 
 ($\sim 18$\AA~difference).
 Furthermore, the He I lines have $v_{cont}$(He I)$=$2000 km s$^{-1}$ while 
 $v_{cont}$(CII)$=$5000 km s$^{-1}$. This means that the C II lines would 
 need to be formed above the helium layer, which would be surprising, if not 
 unphysical, in type Ib SNe. 
 SN 1990I presents therefore good evidence
  for an H$\alpha$ feature. It is not absolutely clearcut however that 
 H$\alpha$ is always responsible for the 
 absorption at about 6300\AA~seen in type Ib SNe. The presence of the 
 H$\beta$ Balmer line would be of great support. 
 Unfortunately the optical depth sufficient to fit the H$\alpha$ trough is so 
 small that the other Balmer features 
 are too weak to be unambiguously detectable ($\tau$(H$\alpha)$ is about 7 
 times greater than $\tau$(H$\beta)$). The second factor that makes H$\beta$ 
 barely discernable is when the contrast velocity of H$\alpha$ is high as
 can be clearly seen from Fig. 1(left).  
 %%%%%%%%%%%%%%%%%%%%%%%%%%%%%%% 
\begin{figure}
 \includegraphics[height=.3\textheight]{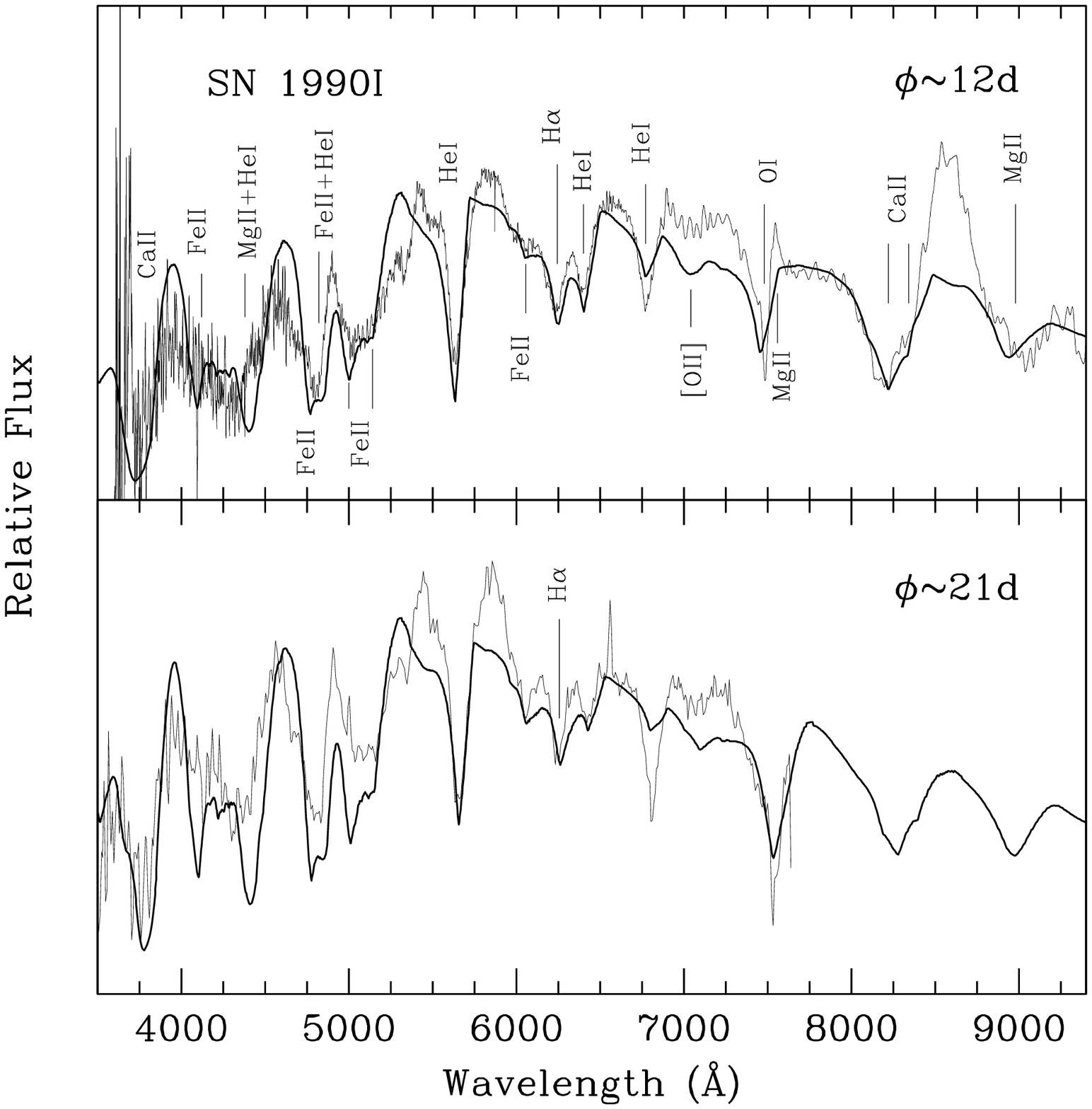}\includegraphics[height=.3\textheight]{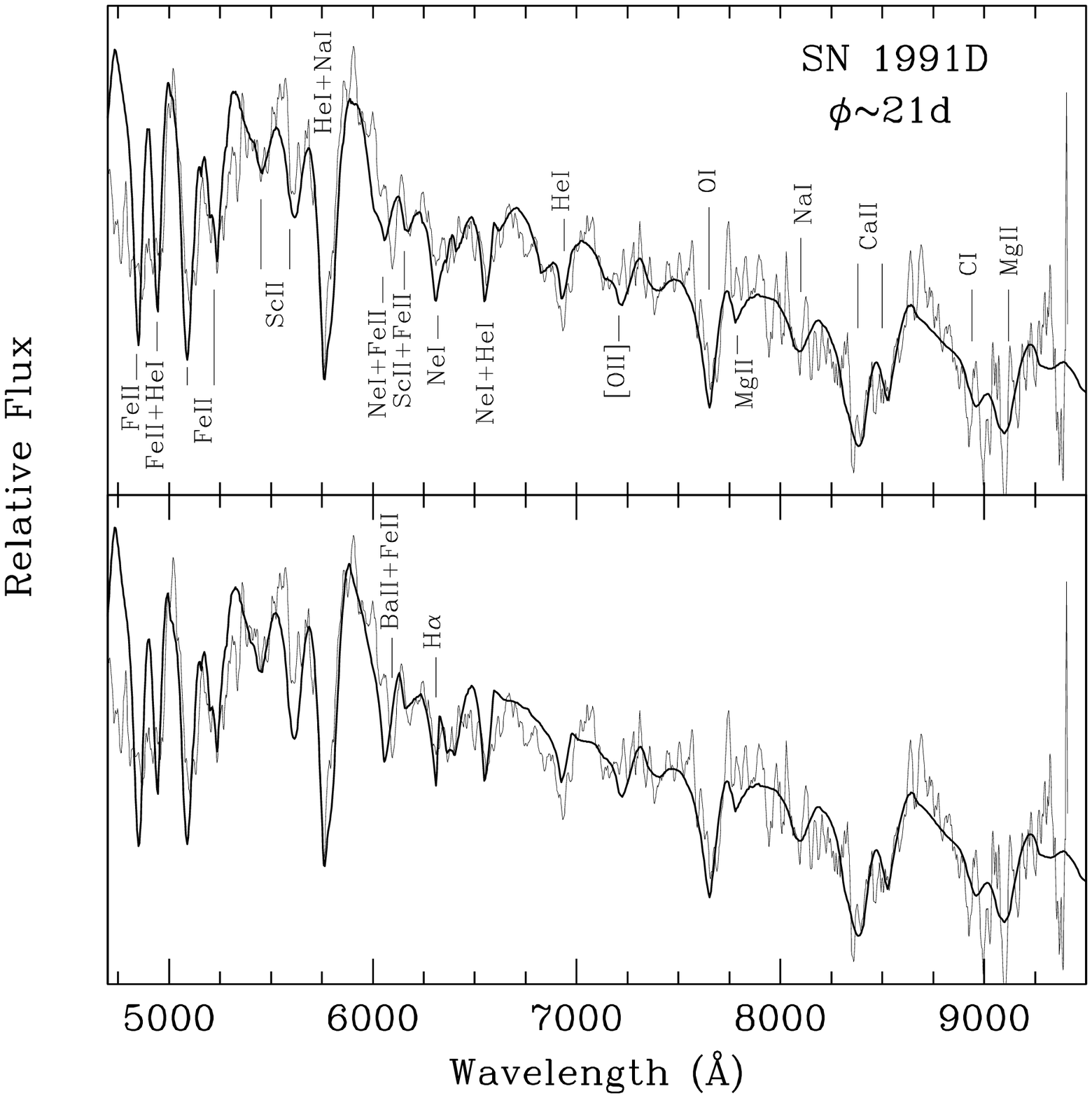}
  \caption{$Left$: SSp fits of SN 1990I spectra at 12d and 21d. $Right$: 
  SSp fit of SN 1991D around 3 weeks. The top panel illustrates the Ne I
  possibility, while H$\alpha$ case is shown in the lower panel.}
\end{figure}
%%%%%%%%%%%%%%%%%%%%%%%%%%%%%%% 
 
 Based on our investigation, the ions that generally might be encountered 
 in shaping the  ``6000$-$6500\AA''~wavelength 
 range are:  H$\alpha$, Ne I, C II, Si II, Sc II, Ca I, He I, Fe II, Si I 
 and Ba II. Here, we propose some selection 
 criteria to decide on the presence of hydrogen for the 6300\AA~feature:\\ 
  $\bf{1-}$ undetached Ne I 6402\AA~line is rejected once its too blue to fit 
 the 6300\AA~trough or/and when the other Ne I lines clearly introduce 
 unwanted features. Similar reasoning applies to the Si II 6355\AA~line.
 $\bf{2-}$ with its contrast velocity as a free parameter, C II 6580\AA~line 
 could be a candidate for the 6300\AA~trough; nevertheless it is ruled out 
 once it exceeds the He I contrast velocity.

 The observed spectra at 12days and 21days are compared in Fig. 2(left) to 
 the synthetic ones with ($v_{phot}$=10000 km s$^{-1}$; T$_{bb}$=5500 K) and 
 ($v_{phot}$=9500 km s$^{-1}$; T$_{bb}$=5400 K), respectively.
 He I is still detached from the photosphere with $v_{cont}$(He I)$=$ 3000
  km s$^{-1}$ at 12d and  $v_{cont}$(He I)$=$ 2500
 km s$^{-1}$ at 21days. The He I 7065\AA~seems to increase in strength 
 relative to the He I lines at 5876\AA~ and 6678\AA, indicating the 
 non-thermal excitation effects are changing but still existent.
 For the ``6000$-$6500\AA''~wavelength region, we find that Fe II, He I and 
 H$\alpha$ are sufficient to 
 reproduce the overall shape in the SSp at both 12days and 21days. We have, 
 however, tested the Ne I, Si II and C II possibilities as candidates for the 
 6250\AA~trough. They are all ruled out by means of our above criteria. We 
 therefore consider H$\alpha$ to be the most likely explanation in SN 1990I.\\

 {\bf SN 1991D :} 
%%%%%%%%%%%%%%%%%%%%%%%%%%
 the event is a peculiar case where its difficult to decide between  H$\alpha$
 and Ne I. The object has narrow features and lower velocities 
 compared to other type Ib events at similar phases \cite{Ben02,Br02}.
 Fig. 2(right) illustrates the fit to the 21d 
 spectrum ($v_{phot} = 4600$ km s$^{-1}$ and  
 T$_{bb}=$7000 K=. The He I lines are evident through our SSp 
 ($v_{cont}$(He I)= 1400 km s$^{-1}$ and $\tau$(He I)=1.8), with the
  Na I D feature ($\tau$(NaI)=4) contributing to the He I 5876\AA~broad 
 P-Cygni profile. The presence of Na I lines is consistent 
 with the good fit around 8100\AA. The undetached Ne I lines possibility
  is illustrated in the top panel ($\tau$(NeI)=2), while the H$\alpha$
 case is tested in the bottom panel. Except for adding Ba II lines in 
 order to help the fit at $\sim$6100\AA, the other ion parameters are kept
  unchanged. The fit to the observed 
 absorption near 6300\AA~ with H$\alpha$, $v_{cont}$(H$\alpha$)= 7400 
 km s$^{-1}$ and $\tau$(H$\alpha$)=0.46, is slightly better compared to 
 the Ne I case. However the SSp, in the bottom panel, does not 
 account for the observed features near 6630\AA~and 
 6840\AA~as does NeI lines in the SSp of the top panel. Ne I remains hence 
 a strong candidate in this type Ib event.
%%%%%%%%%%%%%%%%%%%%%%%%%%%%%%% 
\begin{figure}
  \includegraphics[height=.3\textheight]{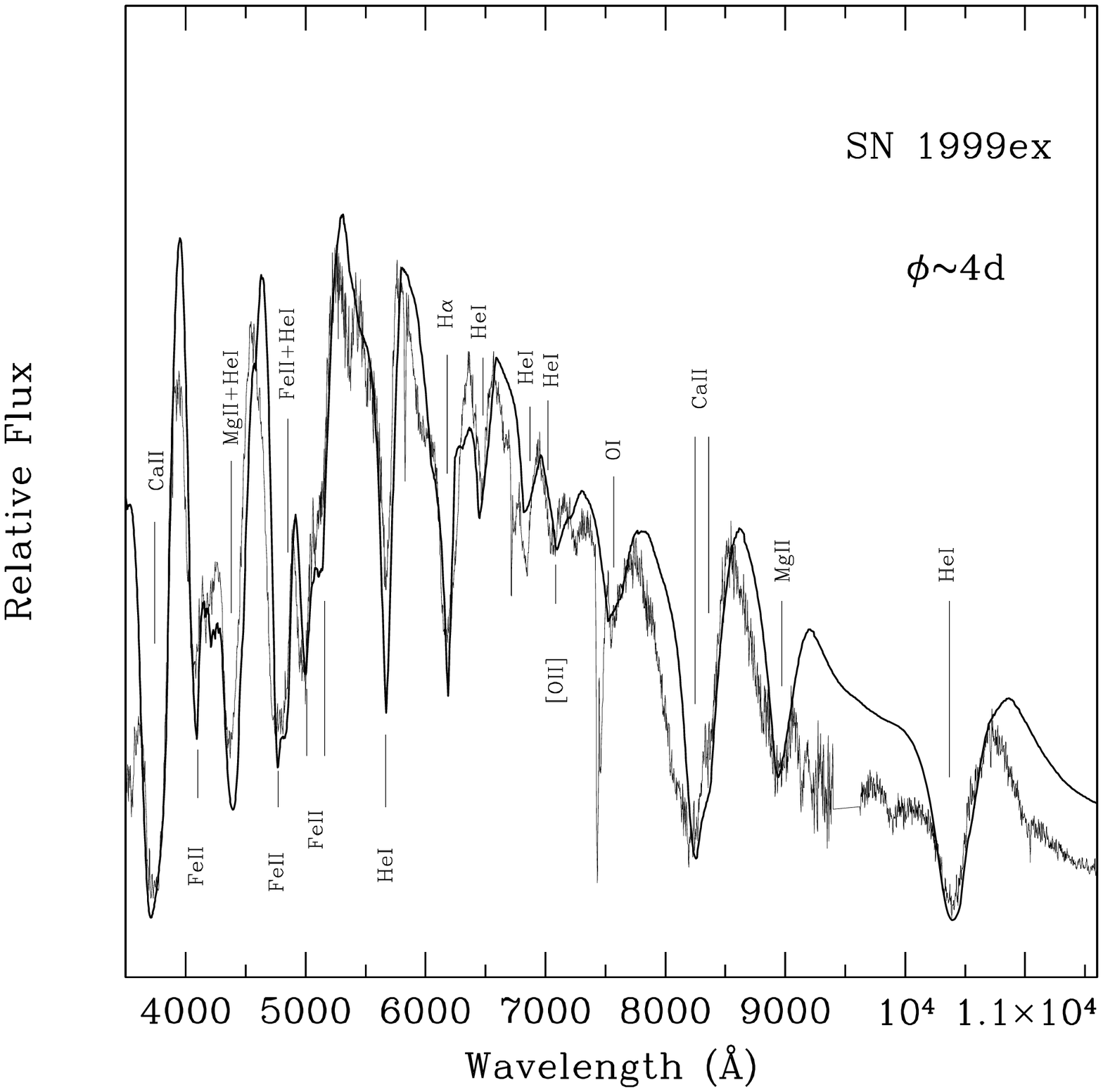}\includegraphics[height=.3\textheight]{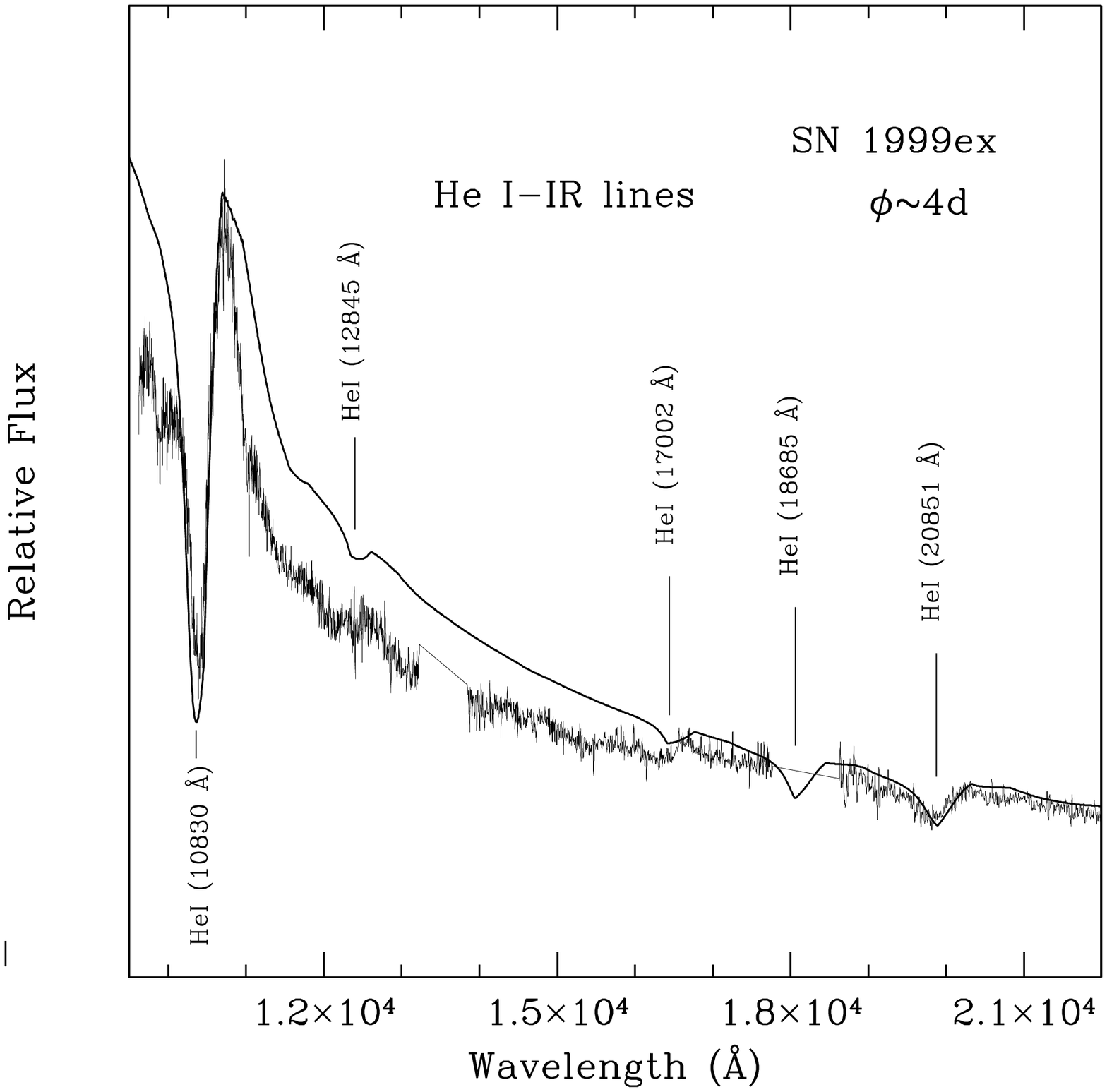}
  \caption{$Left$: SSp fit of SN 1999ex at day 4. Conspicuous line features are
 indicated. $Right$: the 4days IR spectrum compared with SSp that contain only
 lines of He I.}
\end{figure}
%%%%%%%%%%%%%%%%%%%%%%%%%%%%%%% 
\\

{\bf SNe 1999ex:}  
%%%%%%%%%%%%%%%%%%%%%%%%%%
 Because of weak optical He I lines, the object was classified as an 
 intermediate case between Ib and Ic SNe, while the evident trough at 
 $\sim$6250\AA~was attributed to Si II 6355\AA~\cite{Ham02}. 
 
 In Fig. 3(left) the observed 4d spectrum is compared with 
 the best fit that has $v_{phot} = 10000$ km s$^{-1}$ and  T$_{bb}=$5800 K 
 . To check the consistency of He I identification   
  we combine the IR spectrum with the optical one. In fact 
  assigning the following parameters: 
 $v_{cont}$(He I)= 1000 km s$^{-1}$ and $\tau$(He I)=2.35, we obtain a good 
 match to the observed He I profiles including the IR He I 10830\AA. 
 Furthermore, unambiguous evidence for the presence of 
 helium in this event is clear from the extended fit to other 
  He I-IR lines, adopting equal $v_{min}$(He I) and $\tau$(He I) as in the 
 optical part of the spectrum (Fig. 3-right).
  The OI 7773\AA~line with $\tau$=0.5, on the other hand, is found to be not 
 so strong and deep as is the case in most type Ic SNe \cite{Math01}. 
 These two facts point to a Ib nature, more than a Ic class. 
%%%%%%%%%%%%%%%%%%%%%%%%%%%%%%% 
\begin{figure}
  \includegraphics[height=.3\textheight]{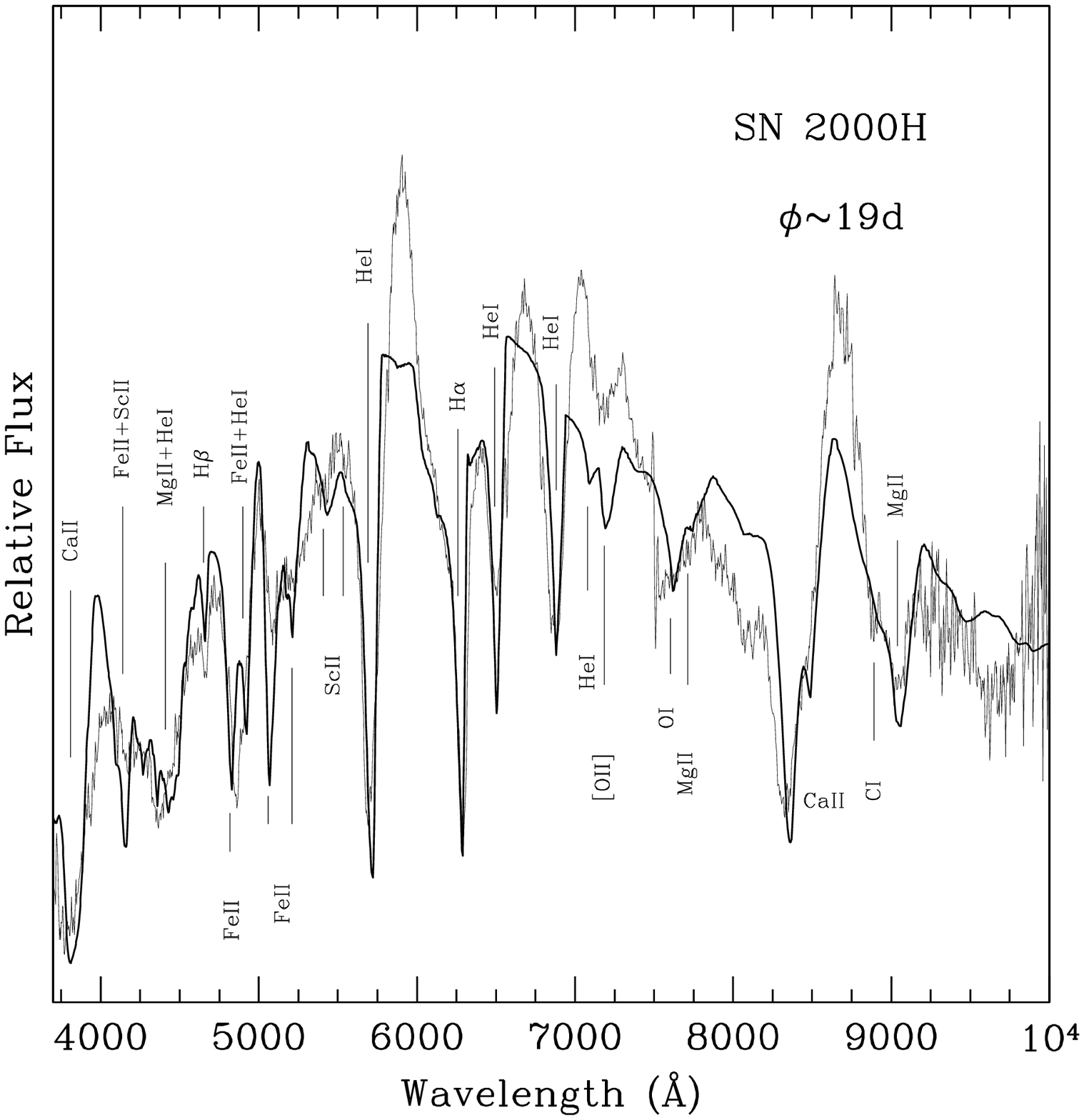}\includegraphics[height=.3\textheight]{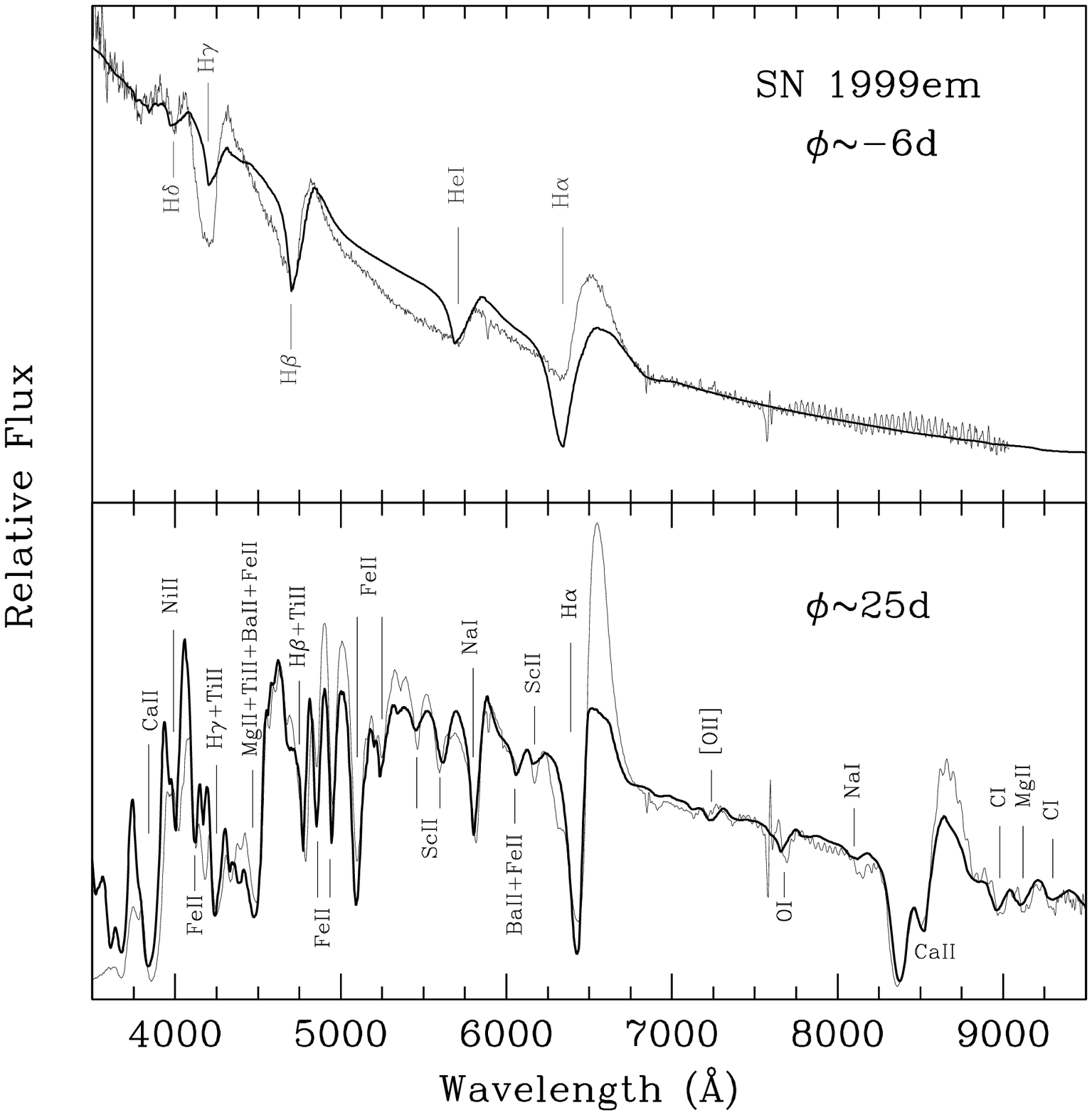}
  \caption{$Left$: SSp fit of SN 2000H at day 19. $Right$: SSp fits
 of SN IIP 1999em observed spectra at day -6(top) and day 25(bottom). The 
 reported phases are normalized to maximum in type Ib-c SNe.}
\end{figure}
%%%%%%%%%%%%%%%%%%%%%%%%%%%%%%% 

 As far as the trough around 6250\AA~is concerned, we checked the Si II 
 identification attributed by Hamuy et al.(2002)\cite{Ham02}.
 We test as well the Ne I possibility. The undetached Ne I fit introduces 
 unwanted features in the spectrum.
 The Si II 6355\AA~is rather blue to account for the feature. We note here 
 that if one adopts C II 6580\AA, then one needs to assign it a very high 
 velocity, about 8000 km s$^{-1}$ higher than the one of He I. 
 The most likely identification therefore remains 
 H$\alpha$. In fact the best fit is achieved using 
 $v_{cont}$(H$\alpha$)= 8000 km s$^{-1}$ and $\tau$(H$\alpha$)=1.45, 
%%%%%%%%%%%%%%%%%%%%%%%%%%%%%%%%%%%%
\\

 {\bf SN 2000H :}
%%%%%%%%%%%%%%%%%%%%%%%%%%%%%%% 
the SN is an example of a type Ib event with a strong and deep trough 
near  6300\AA~\cite{Br02}. 
 Fig. 4(left) compares the 19d spectrum with the best fit SSp ($v_{phot} 
 = 6000$ km s$^{-1}$ and  T$_{bb}=$4600 K).
 Lines of  Ca II, [O II], O I, Mg II are seen to develop. 
 The optical He I lines are noticeably deep 
 ($v_{cont}$(He I)= 2000 km s$^{-1}$ and $\tau$(He I)=5). 
 We note the good match to the $\sim$6250\AA~trough with H$\alpha$, achieved 
 using  $v_{cont}$(H$\alpha$)= 7000 km s$^{-1}$ and $\tau$(H$\alpha$)=2.5. No 
 alternative identification to H$\alpha$, that would be logically acceptable,
  has been found. In addition, the H$\alpha$ 
 identification is supported by the presence of the absorption notch 
 near 4660\AA, well accounted for by H$\beta$ in our resulting synthetic 
 spectra. \\
%%%%%%%%%%%%%%%%%%%%%%%%%%%%%%% 

 {\bf SN IIP 1999em:}
 Fig. 4(right) displays two spectra during the photospheric phase of the 
 typical type II-Plateau SN 1999em \cite{Elm03}. 
 The phases are normalized to maximum light in type Ib-c SNe. The 
 best fit synthetic spectra have $v_{phot} = 10000$ km s$^{-1}$ and 
 T$_{bb}=$10000 K (-6d; upper panel) and  $v_{phot} = 4600$ km s$^{-1}$ 
 and  T$_{bb}=$6400 K (25d; lower panel). At the earliest phase only 
 undetached lines of hydrogen Balmer, He I and a weak contribution from Ca II 
 are sufficient to reproduce the most conspicuous features superimposed on 
 the ``hot'' continuum. The fit with Na I D, 
 at this phase, is poor compared to He I 5876\AA. The He I contribution, 
 in type II SNe, is found to be 
 important shortly after the explosion. In fact in SN 1987A, the 
 He~I 5876\AA~feature was clearly present during 
 the first few days, then rapidly faded and disappeared completely 
 around 1 week after the explosion, when Na I D starts to emerge 
 \cite{Han88}.
 At -6d, Balmer hydrogen features are evident with $\tau \ge 15$ for
 H$\alpha$. The corresponding strong and 
 broad P-Cygni feature cannot be produced completely by the ``SYNOW'' code. 
 This is because in our SSp treatment we are adopting a resonance scattering 
 source function. At day 25, Fig. 4(right-bottom), H$\alpha$ is narrower 
 and has $v_{cont}$= 1000 km s$^{-1}$ and  $\tau=21$. The envelope
 temperature decreases and many lines emerge at this phase. Apart from the 
 hydrogen lines (slightly detached), all the 
 lines are undetached. \\

%%%%%%%%%%%%%%%%%%%%%%%%%%%%%%%%%%
\section{Results and Discussion}
 Based on results presented here and
 in the larger sample study of Elmhamdi et al. (2006)\cite{Elm06}, 
 hydrogen is found to manifest its presence in a different way in the 
 CCSNe sample. SNe of type II and IIb show strong and broad 
 H$\alpha$ P-Cygni emission component, which is absent in type Ib objects.
 Indeed, the H$\alpha$ P-Cygni profile would lose its obvious 
 emission component when it is highly detached. H I Balmer lines have 
 large optical depths, allowing them to be distinctly visible, contrary to 
 normal type Ib SNe, with the exception of some cases with deep and 
 conspicuous H$\alpha$ troughs that present the signature of H$\beta$ as well. 
 Two factors make H$\beta$ barely discernible in type Ib: the optical depth 
 found to fit H$\alpha$ is small and the contrast velocity 
 of H$\alpha$ is high. The opposite is seen in type II and IIb  events.

 In Fig. 5, upper panel, we report the resulting photospheric velocities 
 from our best fits for the whole CCSNe sample \cite{Elm06}. 
 Data for SN 1987A, corresponding to 
 Fe II 5018\AA~absorption, are also shown for comparison (dashed line 
 \cite{Phi88}). The plot indicates the low velocity 
 behaviour of type II SNe, both 1987A and 1999em. SN IIb 1993J follows 
 somewhat similar behaviour as SN Ic 1994I in having higher velocities. 
 Type Ib SNe appear to display a different 
 velocity evolution. The scatter seems to increase at 
 intermediate phases (around 20$-$30 days). This fact can be simply due to 
 the paucity of available observations 
 outside that range. Around day 20, for example, a scatter as high as 
 5000 km s$^{-1}$ is measured. SNe 1990I and 1998dt 
 belong to a class with the higher ``$v_{phot}$'', while objects such as 
 SNe 1991D and 1996aq have the lowest estimated velocities, approaching 
 even type II objects. The remainder of the Ib events 
 follow a similar trend, namely the one described by Branch et
  al. (2002)\cite{Br02}. 

%%%%%%%%%%%%%%%%%%%%%%%%%%%%%%% 
\begin{figure}
\includegraphics[height=.34\textheight]{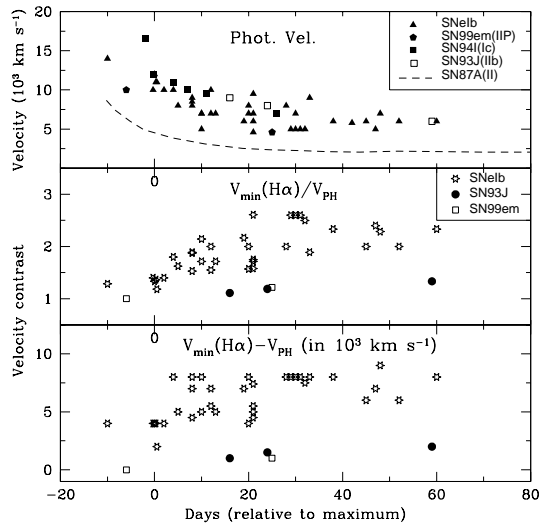}
\caption{The photospheric velocity evolution of the analyzed CCSNe
 (top). The H$\alpha$ contrast velocity evolution is also illustrated (middle 
 and lower panels). }

\end{figure}
%%%%%%%%%%%%%%%%%%%%%%%%%%%%%%% 
 The middle and bottom panels in Fig. 5 display the evolution the 
 of H$\alpha$ contrast velocities. An increasing trend of 
 ``$v_{cont}$(H$\alpha$)'' is found going from type II to IIb to 
 Ib SNe. While in SN IIb 1993J and SN~IIP~1999em the line is found
 to be either undetached or slightly detached, it is 
 highly detached in type Ib events. Moreover, the ``$v_{cont}$(H$\alpha$)'' 
 is found to increase within the first 15 days, 
 reaching values as high as 8000  km s$^{-1}$, and then follows an almost 
 constant evolution. Up to $\sim$60 days, type Ib SNe are found to have 
 hydrogen down to 11000-12000 km s$^{-1}$.
 SNe II appear to have hydrogen down to lower 
 velocities ($\sim$5000 km s$^{-1}$ in SN 1999em).
 In addition, hydrogen in type Ib SNe is found to have very small optical 
 depths independently of the contrast velocity.
 
 An approximate method to estimate the hydrogen mass from the fit 
 results invokes the amount needed to fill a uniform density sphere of 
 radius ``$v \times t$'' at an epoch $t_d$ since explosion.
 The required ion mass can be given by (\cite{Elm06}):
\begin{equation}
 M(M_\odot)\simeq (2.38 \times 10^{-5})~v_4^{3}~ t_d^{2}~\tau(H\alpha);
\end{equation}
 where $t_d$ in days and $v_4$ is in 10$^{4}$ km s$^{-1}$. 
 Although non-thermal excitation and ``NLTE'' effects may be also important 
 for hydrogen, this simple approach seems to give reasonable estimates. For  
 SN Ib 1990I around maximum light a value of 0.02 $M_\odot$ is computed. 
 If we adopt a representative H$\alpha$ optical depth 
 of 0.5 at day 20 with an  H$\alpha$ velocity restriction similar to SN 1990I, 
 the estimated hydrogen mass is of the order 0.015 $M_\odot$. Events with 
 higher velocity widths and/or deeper H$\alpha$ troughs would eject larger 
 amounts. A thin layer of hydrogen, ejected at high velocities down to 
 11000-12000 km s$^{-1}$, appears to be present in almost all the studied
 type Ib events.  
 
 Worth-noticing here is that the presence of H$\alpha$ in type Ic events 
 was recently re-addressed by Branch et al. (2006) \cite{Br06}. A large
 type Ic sample study is clearly needed to assess this issue. However,
 based on these preliminary investigations the progenitor nature of  
 the stripped-envelope SNe should be revised.

 An important by-product results of our fit analysis concerns the behaviour of 
 the O I 7773\AA~line. At intermediate phases, it seems 
 that type Ib objects tend to have low optical depths in this line, while 
 SNe Ic are found to display the deepest profile. SNe IIb an II, at similar 
 phases, are the objects with the lowest O I 7773\AA~optical depth. 
 The stronger and deeper 
 permitted oxygen lines at early phases of SNe Ic spectra might 
 imply that they are less diluted 
 by the presence of a helium envelope. Indeed one might expect oxygen lines to 
 be more prominent for a ``naked'' C/O progenitor core. Despite the paucity of 
 well sampled CCSNe observations, two observational
 aspects tend to reinforce this belief: $\bf{First}$, the forbidden
 lines, especially [OI]6300,~6364\AA, seem to appear
 earlier following a SNe sequence ``Ic$-$Ib$-$IIb$-$II''. In fact the oxygen 
 line emerges at an age of 1-2 months in
 Type Ic SN 1987M \cite{Fil97}. SN Ic 1994I displayed 
 evidence for the line at an age of ~35-50 days \cite{Cloc96b}. 
 While in SN Ib 1990I it was hinted at the 70day spectrum 
 \cite{Elm04}. In other Type Ib SNe it appears
 earlier than in SN 1990I. In SN IIb 1993J, a transition object, the line was
  visible in the 62day spectrum \cite{Barb95}. 
 In SNe II, however, the line appears later: around day 150 in SN 1987A 
 \cite{Cat88} and after day 138 in SN 1992H \cite{Cloc96a}. 
 $\bf{Second}$, the nebular 
 emission line looks decreasing in breadth following the SNe sequence above. 
 More investigations concerning the oxygen issue in CCSNe are under progress
 (Elmhamdi \& Danziger, in preparation).


\begin{thebibliography}{9}
\bibitem{Barb95}
Barbon R., Benetti S., Cappellaro E. et al. 1995, $A\&AS$, {\bf 110}, 513
\bibitem{Ben02}
Benetti S., Branch D., Turatto M. et al. 2002, $MNRAS$, {\bf 336}, 91
\bibitem{Br02}
Branch D., Benetti S., Kasen D. et al. 2002, $ApJ$, {\bf 566}, 1005
\bibitem{Br06}
Branch D., Jeffery D.J., Young T.R. et al. 2006,  \emph{``astro-ph/0604047''}
\bibitem{Cat88}
Catchpole R. M., Whitelock P. A., Feast M. W. et al. 1988, $MNRAS$, 
 {\bf 231}, 75
\bibitem{Cloc96a}
Clocchiatti A., Benetti S., Wheeler J. C. al. 1996a, $AJ$, {\bf 111}, 1286
\bibitem{Cloc96b}
Clocchiatti A., Wheeler J. C., Brotherton M. S. et al. 1996b, $ApJ$, 
 {\bf 462}, 462 
\bibitem{Elm03}
Elmhamdi A., Danziger I. J., Chugai N. N. et al. 2003, $MNRAS$, {\bf 338}, 939
\bibitem{Elm04}
Elmhamdi A., Danziger I. J., Cappellaro E. et al. 2004, $A\&A$, {\bf 426}, 963
\bibitem{Elm06}
Elmhamdi A., Danziger I. J., Branch D. et al. 2006 , $A$\&$A$,{\bf 450}
, 305-330
\bibitem{Fil97}
Filippenko A. V. 1997, $ARA$\&$A$, {\bf 35}, 309 
\bibitem{Ham02}
Hamuy M., Maza J., Pinto P. A. et al. 2002, $AJ$, {\bf 124}, 417
\bibitem{Han88}
Hanuschik R. W. $\&$ Dachs J. 1988, $A\&A$, {\bf 205}, 135
\bibitem{Jeff90}
Jeffery D. J. $\&$ Branch D. 1990, in: J.C. Wheeler, T. Piran, $\&$ S. 
 Weinberg (eds.),``Supernovae'', $Sixth~ Jerusalem~ Winte~ School~ for 
 Theoretical~ Physics$ (Singapore: World Scientific), {\bf 149}
\bibitem{Luc91}
Lucy L. B. 1991, $ApJ$, {\bf 383}, 308
\bibitem{Math01}
Matheson T., Filippenko A. V., Li W., Leonard D. C. $\&$ Shields J. C. 2001, 
 $AJ$, {\bf 121}, 1648
\bibitem{Phi88}
Phillips M. M., Heathcote S. R., Hamuy M. $\&$ Navarrete M. 1988, $AJ$, 
 {\bf 95}, 1087
\end{thebibliography}
\end{document}